\documentclass[epj]{svjour}
\usepackage{graphicx}

\begin{document}



\title{Analysis of $p$Pb collisions at LHC energies in the relativistic diffusion model}

\author{P. Schulz \and G. Wolschin
}                     
%
%
\institute{
Institut f{\"ur} Theoretische Physik
der Universit{\"a}t Heidelberg, 
        Philosophenweg 16,  
        D-69120 Heidelberg, Germany, EU}
\date{Received: date / Revised version: date}

\abstract{
Particle production sources in $p$Pb collisions at LHC energies are investigated in pseudorapidity space as functions of centrality.
The nonequilibrium-statistical relativistic diffusion model (RDM) with three sources is applied
to the analysis of preliminary charged-hadron distributions measured by the ATLAS and ALICE collaborations in  $p$Pb collisions at 5.02 TeV.
The size of the midrapidity source relative to the fragmentation sources in $p$Pb collisions
 is considered as function of 
 centrality, and incident energy. At LHC energies particle production from the mid-rapidity  gluon-gluon source is larger than that from the fragmentation sources at all centralities.
 Conclusions regarding the statistical equilibration process are drawn.\\
} 
\authorrunning{P. Schulz, G. Wolschin}
\titlerunning{Analysis of $p$Pb collisions at LHC energies}
\maketitle



\newpage

\section{Introduction}
Whereas charged-particle production in symmetric heavy-ion collisions has been investigated in great detail
both in the AuAu system at RHIC c.m. energies of 19.6 GeV to 200 GeV \cite{alv11}, and in PbPb at LHC energies of 2.76 TeV per particle pair
\cite{gui13},
asymmetric systems such as $p$Pb at the current LHC c.m.energy of 5.02 TeV are still calling for further exploration.

So far, pseudorapidity distributions of produced charged hadrons in $p$Pb have been measured in a rather limited
range of pseudorapidities $-2<\eta<1.9$ (ALICE collaboration \cite{ab12}) or $|\eta|<2.7$ (ATLAS collaboration \cite{atlas13}),
and centrality-dependent results are presently being assembled \cite{atlas13,toi14}. 

Here we investigate how
these upcoming new results compare with the nonequilibrium-statistical relativistic diffusion model (RDM), which has already proven to
be useful for detailed descriptions and predictions of charged-hadron distributions in symmetric systems at RHIC and LHC energies
\cite{gw13}. Its principal relation to other theoretical models such as the statistical hadronization model \cite{mabe08,pbm95,aa06}, the color glass
condensate \cite{ge10}, or hydrodynamics \cite{alv10} has been discussed in a recent application of the RDM to symmetric systems
\cite{gw15}. The situation for the asymmetric case that is treated here is similar with respect to related models. 

In particular, the mid
rapidity source in the RDM becomes equal to the thermal (equilibrium-statistical) result in the limit time to infinity. In the thermal model --
which has a single source by construction -- there is no limiting fragmentation at LHC energies \cite{cley08}, whereas the RDM with
two additional non equilibrium fragmentation sources does show limiting fragmentation \cite{rgw12} in agreement with the data. 

The longitudinal degrees of freedom are much further away from overall equilibrium than the transverse ones, which show 
equilibrium behaviour at small transverse momenta with collective expansion, and pQCD type events above $p_T \simeq 7$ GeV/$c$ \cite{abe13}.
In this work, we integrate out the transverse motion, and consider only the longitudinal behaviour where the interplay of mid rapidity and
fragmentation sources for particle production becomes obvious.

In the RDM, the shapes of the distribution functions for asymmetric systems are much more sensitive to details of the model
as compared to symmetric systems. Hence, valuable conclusions regarding the statistical equilibration processes
in these relativistic systems can be drawn. Moreover, the calculated results beyond the measured regions in (pseudo-)rapidity space provide detailed predictions for the forthcoming data analyses.

The model is briefly explained in the next section, and in the main part of this note the results are compared with preliminary LHC data.
\section{The model}
The relativistic diffusion model (RDM) has been outlined in \cite{gw04}, and applied to asymmetric systems at RHIC energies in \cite{wobi06}, where dAu has been investigated at 200 GeV. In the linearized version,
the RDM is based on a Fokker-Planck type transport equation for the distribution functions $R_k(y,t)$ in rapidity space, 
\begin{equation}
\frac{\partial}{\partial t} R_{k}(y,t) =
-\frac{1}{\tau_{y}^k}\frac{\partial}
{\partial y}\Bigl[(y_{eq}-y)\cdot R_{k}(y,t)\Bigr]
+D_{y}^{k} \frac{\partial^2}{\partial y^2}
R_{k}(y,t).
\label{fpe}
\end{equation}
\begin{figure}[tph]
\begin{center}
\includegraphics[width=8cm]{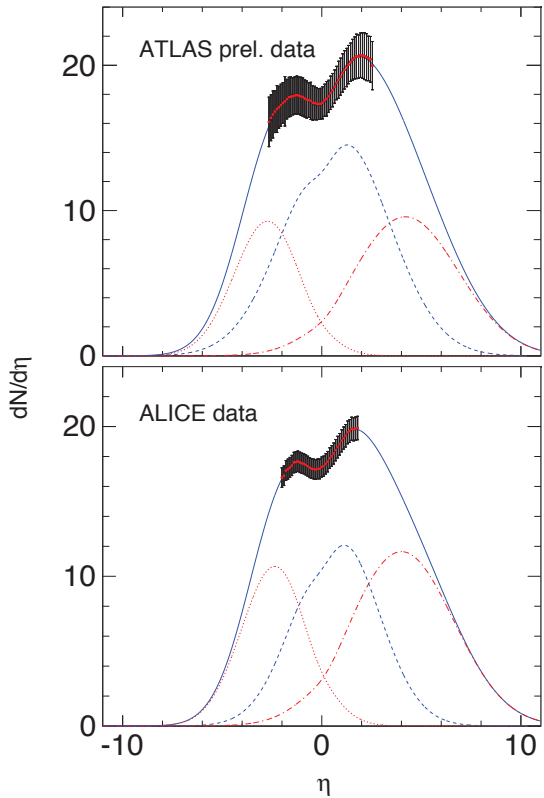}
\caption{\label{fig1}(Color online) The predicted RDM pseudorapidity distribution functions for charged hadrons in minimum bias $p$Pb collisions
at LHC c.m. energy of 5.02 TeV shown here are $\chi^2$-optimized in the mid-rapidity region to the preliminary ATLAS (top) \cite{atlas13}, and published ALICE (bottom) data  \cite{ab12} (systematic error bars only for ALICE). Corresponding $\chi^2$-values per degrees of freedom are 0.436/45  for ATLAS, and 0.388/31 for the ALICE data. The measured preliminary ATLAS data (0--90\%) have been scaled by the geometric cross-section ratio $\sigma(0-90\%)/\sigma(0-100\%)\simeq 0.898$ to make them directly comparable to the minimum-bias ALICE results.
The underlying distributions in the three--sources RDM are also shown.}
\end{center}
\end{figure}

The total distribution function comprises an incoherent superposition of three sources:
The source $R_1$ represents the fragmentation distribution arising from interactions of valence quarks in the projectile (proton) with low-$x$ gluons in the target (Pb), $R_2$ the corresponding fragmentation distribution due to interactions of valence quarks in the target with gluons in the projectile, and $R_3 = R_{gg}$ the midrapidity distribution function arising from the mutual interaction of -- mostly -- low-$x$ gluons in the fireball. The diffusion coefficients $D_y^k$ account for the broadening of the distribution functions due to particle production and collisions, whereas the drift terms cause shifts of the distribution functions towards the equilibrium value, and constrain the diffusion. The  midrapidity source is already centered at the centrality-dependent equilibrium value of the rapidity,
\begin{equation}
y_{eq}(b)=-\frac{1}{2}\ln{\frac{\langle m_1^{T}(b)\rangle\exp(y_{max})+\langle m_2^{T}(b)\rangle\exp(-y_{max})}
{\langle m_2^{T}(b)\rangle\exp(y_{max})+\langle m_1^{T}(b)\rangle\exp(-y_{max})}}
\label{eq}
\end{equation}
with the beam rapidities $y_{beam}=\mp y_{max}=\mp\ln(\sqrt{s_{NN}}/m_p)$, the average transverse masses $\langle m_{1,2}^T(b)\rangle =
\sqrt{m^2_{1,2}(b)+\langle p_T\rangle^2}$, and participant masses $m_{1,2}(b)$ of the $p$- and Pb-like participants in $p$Pb collisions that depend on the impact parameter $b$. 
For very
 high energy such that $\exp{(-y_{max})}
\simeq 0$ the equilibrium value of the rapidity becomes
\begin{equation}
y_{eq}(b)\simeq 0.5 \ln {\frac{\langle m_2^{T}(b)\rangle}{\langle m_1^{T}(b)\rangle}}.
\label{eq1}
\end{equation}
For sufficiently large times $t\to \infty$ -- in statistical equilibrium -- all three distribution functions tend to be centered at $y_{eq}(b)$. The relevant quantity that determines whether the centers of the fragmentation distributions reach $y_{eq}$ is the ratio of the centrality-dependent interaction time, and the source-dependent rapidity relaxation times, $t_{int}/\tau_{y}^{1,2}$, since for asymmetric systems
\begin{equation}
\langle y_{1,2}(t)\rangle=y_{eq}[1-\exp(-t/\tau_{y}^{1,2})] \mp y_{max}\exp{(-t/\tau_{y}^{1,2})}.
\label{mean}
\end{equation}

\begin{table*}
\caption{RDM parameters (mean values $<y_{1,2}>$ and widths $\Gamma_i$), and particle content of the sources for centrality-dependent $p$Pb collisions at 5.02 TeV. See text for the number of RDM parameters. This $\chi^2$-optimization (per numbers of degrees of freedom, dof) uses ROOT \cite{root97}, preliminary ALICE charged-hadron $dN/d\eta$ data \cite{toi14}, and
equilibrium values $y_{eq}$ of the mid-rapidity source. 
The mean transverse momenta $\langle p_T \rangle$ are taken from the data \cite{ab13}.
The beam rapidity is $y_{beam}=\mp y_{max}= \mp 8.586.$}
\label{tab1}
\begin{center}
\begin{tabular}{rccccccccrrr}
\hline
centrality&$\langle p_T \rangle$&$y_{eq}$&$\langle y_1\rangle$&$\langle y_2\rangle$&$\Gamma_1$&$\Gamma_2$&$\Gamma_{gg}$&$N_1$&$N_2$&$N_{gg}$&$\chi^2/$dof\\
(\%)&GeV/$c$\\
\hline
0--5&0.800&0.887&-2.050&2.900&4.819&4.500&7.500&100&119&308&7.551 / 32\\
5--10&0.779&0.886&-2.090&3.005&4.591&5.000&7.994&75&104&248&7.967 / 33\\
10--20&0.767&0.876&-2.100&3.100&4.500&5.100&8.750&53&80&236&4.447 / 33\\
20--40&0.743&0.818&-2.300&3.300&4.353&5.150&8.900&32&47&200&1.461 / 32\\
40--60&0.713&0.817&-2.750&4.500&4.350&6.000&9.500&18&31&157&0.693 / 33\\
60--80&0.660&0.563&-2.760&4.700&4.200&6.500&10.000&9&18&95&0.463 / 33\\
80--100&0.608&0.129&-2.800&4.710&4.101&6.998&10.050&4&9&31&0.137 / 32\\
\hline
\end{tabular}
\end{center}
\end{table*}

It has turned out \cite{gw13} that at RHIC and LHC energies, the interaction time in symmetric systems is too small for the system to reach statistical equilibrium \cite{gw13}. It is only the central source that reaches equilibrium in rapidity space -- although in transverse momentum space, the distribution functions of charged hadrons are actually very close to thermal equilibrium. The mean value of the moving mid-rapidity source remains $\langle y_3\rangle=y_{eq}(b)$ at all times, and the variance approaches equilibrium according to 
$\sigma_{k}^{2}(t)=D_{y}^{k}\tau_{y}^k[1-\exp(-2t/\tau_{y}^k)]$, for $\delta-$function initial conditions as in \cite{gw13}. 

The model has four parameters in case of symmetric
systems: the ratio of interaction time and relaxation time $t_{int}/\tau_{y}^{1=2}$ (or the corresponding mean value as obtained from Eq.~\ref{mean}), the variances $\sigma_{1=2}^{2}$ (or FWHM $\Gamma_{1=2}$) of the fragmentation sources and $\sigma_{gg}^{2}$ (or FWHM $\Gamma_{gg}$) of the mid rapidity source, plus the ratio of particles produced in the mid rapidity source, and in total for a given centrality. (If the total number of produced charged hadrons is not considered as experimental input, but as an additional parameter, we have five parameters). These parameters are determined in $\chi^2-$ minimizations of the analytical solutions with respect to the data.

In case of asymmetric systems that are investigated in this work, the total distribution function is even more sensitive to the precise shape and the interplay of the three sub-distribution functions, together with the Jacobian transformation between rapidity and pseudorapidity space. As a particular example, the shoulders of the distribution functions are not symmetric: The fall-off  depends on centrality, and is predicted to be steeper on the $p$-like side than on the Pb-like side, in particular in central collisions. Whereas this is indeed visible in dAu data at RHIC energies \cite{alv11,wobi06}, LHC $p$Pb data do not yet reach the region in (pseudo-)rapidity space where this specific model prediction can be tested.

For asymmetric systems, the mean value of the mid rapidity distribution is calculated as described above, but three additional parameters enter the calculation because the rapidity relaxation times and hence, the mean values of the fragmentation distributions as well as the widths of all three partial distributions and the particle numbers in the fragmentation sources differ from each other. If one takes the number of produced charged hadrons in the fragmentation sources to be proportional to the corresponding numbers of participants as proposed in 
\cite{wobi06}, one needs only two additional parameters for asymmetric systems.

Since particle identification is not available for $p$Pb hadron rapidity distributions, the calculated rapidity density distribution functions $dN/dy$ have to be converted to pseudorapidity space, $\eta=-$ln[tan($\theta / 2)]$,  before they can be compared to data so that the RDM-parameters can be deduced in $\chi^2-$minimizations.  The Jacobian $dy/d\eta=
\cosh({\eta})[1+(m/ p_{T})^{2}+\sinh^{2}(\eta)]^{-1/2}$ that mediates the conversion depends not only on the masses of the particles, but also on the transverse momentum, and a careful consideration of both as detailed in \cite{rgw12} is required. The effect of the transformation is most pronounced at small transverse momenta, and in the mid rapidity range.

In particular, it is not sufficient to consider only the mean transverse momentum, but rather precise results can be obtained by using the pion mass $m_{\pi}$, and calculating an effective transverse momentum $\langle p_T^{eff}\rangle$ such that the experimentally determined Jacobian $J_{{y=0}}$ of the total charged-hadron distribution at $y=0$ is exactly reproduced. This yields
$\langle p_T^{eff}\rangle(b)=m_{\pi}J_{y=0}/\sqrt{1-J_{y=0}^2(b)}$. 

The effective transverse momenta are considerably smaller than the mean transverse momenta determined from the $p_T-$distributions (Table~\ref{tab1}), and the corresponding effect on the Jacobian is larger than what would be obtained with $\langle p_T\rangle$ from the transverse momentum distributions for each particle species. For PbPb at 2.76 TeV we had obtained in 0-5\% central collisions $\langle p_T^{eff}\rangle \simeq   0.323 \langle p_T\rangle$. The influence of the Jacobian on $dN/d\eta$ at LHC energies remains, however, essentially confined to the midrapidty source, its effect on the fragmentation sources is marginal.

Whereas the Jacobian produces a mid-rapidity minimum in charged-hadron pseudorapidity distributions for symmetric systems at LHC energies, its effect is not sufficient to generate the dip seen in the PbPb data at 2.76 TeV, which is indeed more pronounced than the one in AuAu at RHIC energies -- although the contribution due to the Jacobian decreases with increasing energy. It has been shown in \cite{gw13} that an understanding of the mid-rapidity minimum requires both, the proper consideration of the Jacobian, plus the interplay of the three sources. In particular, the fragmentation sources move further apart at higher energies, such that the gluon--gluon source determines the mid-rapidity yield almost exclusively.

In asymmetric systems like $p$Pb at 5.02 TeV, the effect of the Jacobian is by itself not strong enough to produce a local minimum in the mid-rapidity source, see Fig.~\ref{fig1} for the minimum-bias pseudorapidity distribution of charged hadrons. Here we have used the same relation between $\langle p_T^{eff}\rangle$ and $\langle p_T\rangle$ as for PbPb. The Jacobian causes a deformation of the mid rapidity source, and the local minimum seen in data emerges when the gluon--gluon source is added incoherently to the two asymmetric fragmentation sources to yield the measured pseudorapidity density distribution function. 

It is interesting to observe that the fragmentation sources become directly visible in measurements (SPS, RHIC) or predictions (LHC) of net-baryon or net-proton (proton minus antiproton) rapidity distributions \cite{mtw09}, where the mid-rapidity gluon-gluon source cancels out. In a recent calculation of the net-proton fragmentation sources that builds upon and extends the QCD-based model developed in \cite{mtw09}, results for 5.02 TeV $p$Pb have been obtained \cite{dur14}. Whereas the net-proton yields are probably too small to be measurable, the charged-hadron yields from these sources are evidently substantial, and decisive for a proper understanding of the pseudorapidity distributions of produced charged hadrons.

\section{Results compared with LHC data}

For minimum-bias (0--100\%) $p$Pb collisions at a c.m. energy of $\sqrt{s_{NN}} = 5.02$ TeV, the ALICE collaboration has already published data \cite{ab12}, which have been analysed within the RDM in \cite{gw13}, see also the lower frame in Fig.~\ref{fig1}.
Meanwhile the ATLAS collaboration has also presented preliminary 0--90\% minimum-bias data in \cite{atlas13}, which we analyse accordingly in the upper frame of Fig.~\ref{fig1}. Here we have scaled the measured 0--90\% ATLAS results by the geometrical cross-section ratio $\sigma(0-90\%)/\sigma(0-100\%)\simeq 0.898$ to make them directly comparable to the minimum-bias ALICE results. 
\begin{figure}[tph]
\begin{center}
\includegraphics[width=9cm]{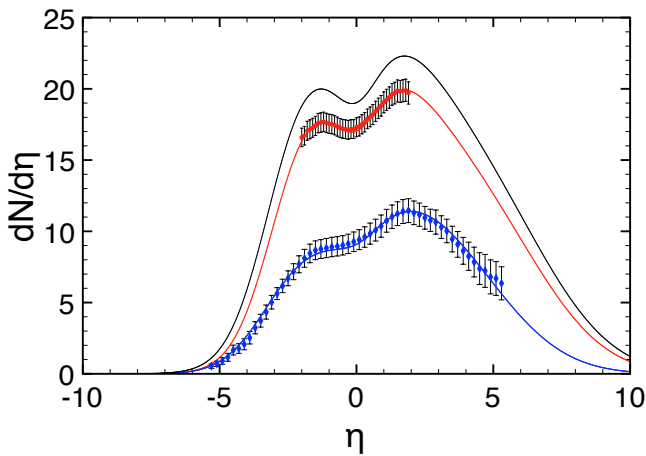}
\caption{\label{fig2}(Color online) The predicted RDM pseudorapidity density distribution functions for charged hadrons in minimum-bias $p$Pb collisions
at the future LHC c.m. energy of 8.16 TeV (upper solid curve) is shown together with the result at 5.02 TeV (middle curve with ALICE data \cite{ab12}, see \cite{gw13} and lower frame in Fig.~\ref{fig1}), and the minimum-bias distribution function in dAu at 0.2 TeV \cite{bb04} with the RDM result from \cite{wobi06}, lower curve.}
\end{center}
\end{figure}
\begin{figure}[tph]
\begin{center}
\includegraphics[width=9cm]{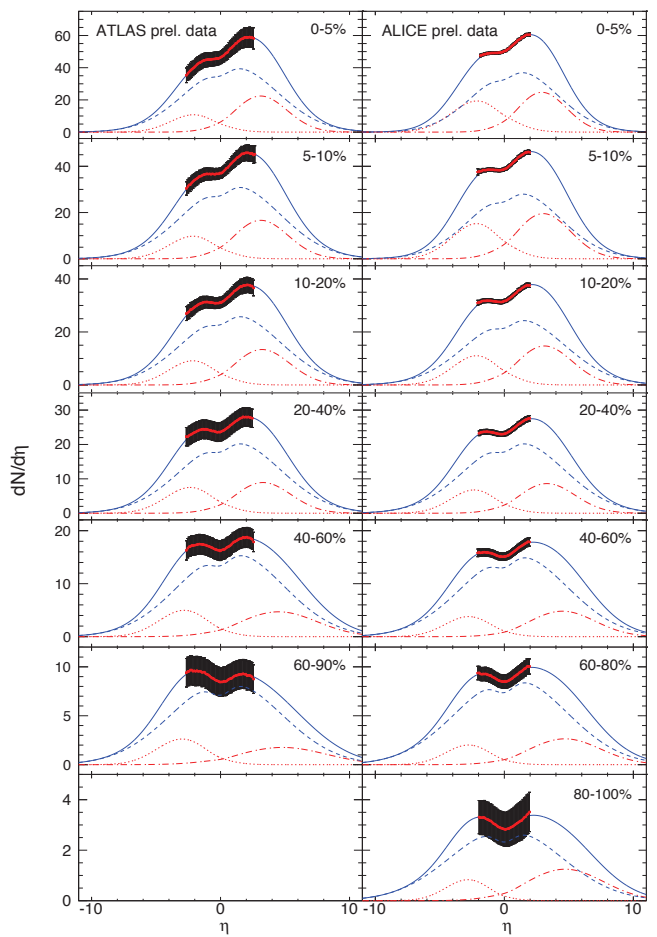}
\caption{\label{fig3}(Color online) The centrality-dependent RDM pseudorapidity distribution functions for charged hadrons in $p$Pb collisions
at LHC c.m. energy of 5.02 TeV are adjusted in the mid-rapidity region through $\chi^2-$minimizations to the preliminary ATLAS (left) \cite{atlas13} and ALICE (right)  \cite{toi14} data  (systematic error bars only for ALICE).
The underlying distributions in the three--sources RDM are also shown, with the dashed curves
arising from gluon-gluon collisions, the dash-dotted curves from valence quark-gluon events in the Pb-like region, and the dotted curves in the proton-like region
(fragmentation sources).}
\end{center}
\end{figure}

As is obvious from the figure, the ATLAS and ALICE results are compatible with each other within the systematic error bars. Using the current version of the object-oriented data analysis framework ROOT \cite{root97}, the RDM-parameters are determined through a  
$\chi^2-$minimization of the difference between the analytical solutions of Eq.~\ref{fpe} and measured $dN/d\eta$ distributions. This yields, however, a larger mid-rapidity source for the preliminary ATLAS results (49 \% of the total particle content), than for the ALICE data (35\%). In view of the rather restricted range of the data in $\eta-$space, 
we expect that these discrepancies vanish once more precise final data in a larger $\eta-$range become available. 

With these forthcoming data, it will be very interesting to compare the predicted steeper slope on the proton-like side with the less steep one on the Pb-like side. A qualitatively similar behaviour had been observed in dAu collisions at RHIC energies of 200 GeV by PHOBOS \cite{alv11}. 

At RHIC energy the mid-rapidity source in minimum-bias dAu carried only 23\% of the total particle content \cite{wobi06}. This is in accordance with the expectation that the gluon-gluon moving particle production source that is almost in statistical equilibrium, and centered at low rapidities $y_{eq}$ in the fireball, becomes more important at LHC energies. 

Future LHC data on asymmetric systems will include charged-hadron production in $p$Pb collisions at a c.m. energy
of $\sqrt{s_{pN}} = \sqrt{Z_1Z_2/(A_1A_2)}\cdot 2 p_p = 8.16$ TeV for a proton momentum of $p_p = 6.5$ TeV/$c$. An RDM prediction for minimum-bias results at this energy is shown in Fig.~\ref{fig2}. 
Here the particle content in the three sources, the corresponding mean values and widths of the distribution functions have been extrapolated based on the low-energy minimum-bias values for dAu at 200 GeV \cite{bb04,wobi06}, and the results for $p$Pb at 5.02 TeV \cite{gw13}. 

Since the energy gap between these two cases is very large, the extrapolation remains somewhat uncertain. In particular, the particle content and the widths may turn out to differ slightly from what is shown in Fig~\ref{fig2}, where the increase in the total yield -- integrated over pseudorapidity -- is 
rather small: The total charged-hadron content in minimum-bias collisions is $N_{tot}$ = 210 at 8.16 TeV, compared to 181 at 5.02 TeV.
Note, however, that the difference will be more pronounced in central collisions.

Of particular interest is the investigation of the centrality dependence, which has been carried out experimentally by both the ATLAS \cite{atlas13} and ALICE \cite{toi14}
collaborations. Several centrality bins are identical and can be compared directly, with the following exceptions: For 0-5\% the ATLAS values are deduced from the 0-1\% and 1-5\% preliminary data \cite{atlas13} by weighting with the geometrical cross sections, similarly for 20-40\%, which is deduced from the 20-30\% and 30-40\% preliminary ATLAS data (here the geometrical cross sections are identical for the two bins). No ATLAS data are available for 90-100\%. For ALICE, preliminary data for various different centrality estimators are available. Here we have chosen the results from the estimator that orders the events based on the number of clusters measured in the second layer of the Silicon Pixel Detector (CL1, \cite{toi14}). 
\begin{figure}[tph]
\begin{center}
\includegraphics[width=9cm]{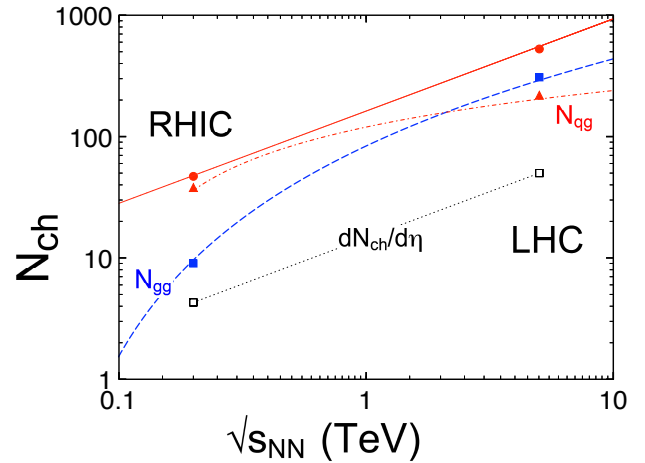}
\caption{\label{fig4}(Color online)Produced charged particles in central dAu collisions at $\sqrt{s_{NN}} = 200$ GeV (RHIC/ PHOBOS data \cite {alv11}; see also \cite{wobi06}) and 0-5 \% pPb at
$\sqrt{s_{pN}} = 5.02$ TeV (LHC/ ALICE data \cite {toi14}). The upper solid line is a power-law fit to the data with the RHIC 0-20 \% results scaled to 0-5\% according to the cross section, and $N_{tot}\propto s_{pN}^{0.38}$. The number of charged particles produced in the fragmentation sources (triangles and dash-dotted curve) is $N_{qg}\propto \ln(s_{pN})$,
and the number of charged hadrons produced in the mid rapidity gluon-gluon source (squares and dashed curve) is $N_{gg}\propto \ln^3(s_{pN})$ (see text for details). The value of the pseudorapidity density for 0-5\% central collisions at the equilibrium value of $\eta$ is also indicated, open squares and dotted line.}
\end{center}
\end{figure}

The results from the $\chi^2-$optimization of the RDM-parameters for both centrality-dependent data sets are shown in Fig~\ref{fig2}, and Table~\ref{tab1} (for ALICE). 
As an additional constraint to the optimisation procedure we require the RDM-parameters (mean values $<y_{1,2}>$ and widths $\Gamma_{1,2,gg}$) to be monotonic functions of centrality, by choosing appropriate initial conditions.
The consistency between the two preliminary data sets, and the corresponding RDM parameters is quite satisfactory. A small problem arises in the 60-90\% centrality bin, where the ATLAS $p$-like maximum is above the Pb-like yield, but this deviation is within the systematic error bars. 

At all centralities, the mid-rapidity gluon-gluon source (dashed curves in Fig~\ref{fig3}) has a larger particle content than the sum of the two fragmentation sources, with a relative charged-hadron content rising monotonically from 58\% in the 0--5\% centrality bin (referring to the ALICE preliminary data) to 78\% in the 60-80\% centrality bin. In contrast, at a RHIC energy of 0.2 TeV, the mid-rapidity source in dAu had been found to have a larger particle content than the sum of the fragmentation sources only for 60--80\% centrality \cite{wobi06} and more peripheral collisions, where the valence-quark content is small and hence, the relative contribution due to the fragmentation sources less important.

For all three sources, the particle contents decrease monotonically towards more peripheral collisions, whereas the fluctuations, and the associated values of the full widths at half maximum $\Gamma_{2,gg} = \sqrt{(8\ln2)}\sigma_{2,gg}$ increase monotonically. The width $\Gamma_{1}$ of the proton-like source decreases slightly for more peripheral collisions. These trends are essentially in line with what we had found earlier for dAu at RHIC energies \cite{wobi06}. 

For the centrality-dependent mean values of the fragmentation sources, we have listed in Table~\ref{tab1} $\langle y_{1,2} \rangle$ rather than the RDM-parameters $t_{int}/\tau_y^{1,2}$ since these can be calculated directly from Eq.~(\ref{mean}). Whereas the centrality dependence of the mean value of the proton-like distribution function is rather weak, a considerable drift in the mean value of the Pb-like distribution function is observed when the collisions become more central, and the interaction time $t_{int}$ increases. 

The mid-rapidity distribution is always centered at the
impact-parameter dependent calculated values of $y_{eq}$, which are also listed in Table~\ref{tab1}.
Obviously the centrality-dependent shapes of the total distribution functions are very sensitive to the precise values of the drift and also of $y_{eq}$, which are properly described in the nonequilibrium-statistical RDM with a linearised drift term.

Of particular interest is the evolution of the particle content in the three sources as function of the c.m. energy. For the PbPb and AuAu systems, this has been done in \cite{gw13} based on four RHIC central-collision data points between 19.6 and 200 GeV, and the LHC point at 2.76 TeV. It has been found that the total charged-hadron production is following a power law $N_{tot }\propto (s_{NN}/s_0)^{0.23}$, whereas the particle content in the fragmentation sources is $Nqg \propto \ln{(s_{NN}/s_0})$. The particle content in the mid-rapidity source is not too far from a power law in the intermediate energy range 0.1--2.76 TeV \cite{gw13}, but closer inspection reveals that it obeys $N_{gg} \propto \ln^3{(s_{NN}/s_0)}$
\cite{gw15}.

For a corresponding investigation in the asymmetric case, we presently have only two values, namely 200 GeV dAu and 5.02 TeV $p$Pb. A double-logarithmic plot for the particle content in 0-5\% central collisions is shown in Fig~\ref{fig4}. The total particle content is shown as a power-law, $N_{tot} = 4.9\cdot (s_{pN}/s_0)^{0.38} $ with $s_0 = 10^2$ GeV$^2$.
The particle content in both fragmentation sources is $Nqg = 26 \ln{(s_{pN}/s_0}) $ with $s_0 = 10^4$ GeV$^2$. The charged-hadron content in the gluon-gluon source is $ N_{gg} =  0.34 \ln^3{(s_{pN}/s_0)} $ with $s_0 = 1.9\cdot 10^3 $ GeV$^2$.
To assess how accurate these dependencies are, more data at different energies would be needed.
\section{Conclusions}
The analysis of charged-hadron pseudorapidity distributions in 5.02 TeV $p$Pb collisions in a nonequilibrium-statistical three-sources relativistic diffusion model reveals the charged-hadron content in the fragmentation- and mid-rapidity sources for particle production. We have determined the RDM-parameters in $\chi^2-$
minimizations with respect to preliminary ALICE and ATLAS data. At all centralities, the mid-rapidity source has the largest particle content, but the fragmentation sources are necessary for a detailed understanding of the centrality-dependent shape of the total distribution functions. It is the interplay of the three sub-distributions together with the effect of the Jacobian transformation that determines the pseudorapidity density distribution of produced charged hadrons. For minimum-bias collisions, we have performed a prediction for 8.16 TeV $p$Pb that can be tested in the forthcoming LHC experiment.

The shapes of the total distribution functions indicate that the system has not reached statistical equilibrium. In particular, the centres of the fragmentation distributions remain far from the equilibrium values $y_{eq}(b)$: It is not just the produced-particle yields that determine whether the emitting source is in statistical equilibrium, but rather the shapes of the distribution functions. A clarifying analogy may be found in the distribution function of the cosmic microwave background radiation, which -- apart from the fluctuations -- is indeed in  perfect thermal equilibrium. This is, however, not the case for particle production in relativistic heavy-ion collisions when the full phase space is taken into consideration -- although local equilibrium in the hydrodynamic sense appears to be achieved for events with $p_T \le 6-8$ GeV/$c$ as may be inferred from the success of hydrodynamics to describe bulk properties.





\vspace{.4cm}
\bf{Acknowledgments}\\\\
\rm
We are grateful to the ALICE Collaboration for making their preliminary centrality-dependent $p$Pb results  \cite{toi14} available.
The ATLAS results have been read off the graph in \cite{atlas13}.
This work has partially been supported by the ExtreMe Matter Institute EMMI.\\\\
\rm
\bibliographystyle{epjc}
\bibliography{gw_epja_nt}
\end{document}